\documentclass[aps, prd, amsmath, twocolumn, superscriptaddress, preprintnumbers]{revtex4-1}
\usepackage{amssymb, esvect, amsmath, graphicx, latexsym, amsthm, slashed, hyperref, color, dblfloatfix, placeins}
 \DeclareMathOperator{\kev}{KeV} \DeclareMathOperator{\mev}{MeV}        \DeclareMathOperator{\s}{s}     
\newcommand{\cA}{{\cal A}}     \newcommand{\cL}{{\cal L}} \newcommand{\cM}{{\cal M}}  \newcommand{\cO}{{\cal O}}   
\newcommand{\ep}{\epsilon}  
  
\newcommand{\pt}{\partial}   \def\oL{\overline}
\newcommand{\pL}{\left(} \newcommand{\pR}{\right)} \newcommand{\bL}{\left[} \newcommand{\bR}{\right]}   \newcommand{\mL}{\left|} \newcommand{\mR}{\right|}
\newcommand{\beq}{\begin{equation}} \newcommand{\eeq}{\end{equation}}
\newcommand{\bea}{\begin{eqnarray}} \newcommand{\eea}{\end{eqnarray}}
\newcommand{\alg}[1]{\begin{align} \begin{split} #1 \end{split}  \end{align}}

\newcommand{\Eq}[1]{Eq.~(\ref{#1})} \newcommand{\Eqs}[2]{Eqs.~(\ref{#1}) and (\ref{#2})} 
  
\newcommand{\Fig}[1]{Fig.~\ref{#1}} 
\newcommand{\Tab}[1]{Tab.~\ref{#1}}

\newcommand{\EH}[1]{\text{EH, #1}} \newcommand{\eh}{\text{EH}}
\DeclareMathOperator{\tr}{Tr}

\begin{document}
\preprint{ACFI-T17-08, YITP-SB-17-14}

\title{Dark Photon Decay Beyond The Euler-Heisenberg Limit}
\author{Samuel D.~McDermott}
\affiliation{C.~N.~Yang Institute for Theoretical Physics, Stony Brook, NY 11794}
\author{Hiren H.~Patel}
\affiliation{Amherst Center for Fundamental Interactions, Department of Physics, University of Massachusetts, Amherst, MA 01003}
\author{Harikrishnan Ramani}
\affiliation{C.~N.~Yang Institute for Theoretical Physics, Stony Brook, NY 11794}
\date{\today}

\begin{abstract}
We calculate the exact width for a dark photon decaying to three photons at one loop order for dark photon masses $m'$ below the $e^+e^-$ production threshold of $2m_e$. We find substantial deviations from previous results derived from the lowest order Euler-Heisenberg effective Lagrangian in the range $m_e \lesssim m' \leq 2m_e$, where higher order terms in the derivative expansion are nonnegligible. This mass range is precisely where the three photon decay takes place on cosmologically relevant timescales.  Our improved analysis opens a window for dark photons in the range $850\kev \lesssim m' \leq 2m_e$, $10^{-5} \lesssim \ep \lesssim 10^{-4}$.
\end{abstract}

\maketitle

\section{Introduction}

Matter with no Standard Model gauge charges may be charged under an abelian gauge group of its own. The corresponding ``dark photon'' is an ingredient of many compelling extensions to the Standard Model \cite{Essig:2013lka, Alexander:2016aln}. A new gauge boson of this type may kinetically mix with the Standard Model photon \cite{Holdom:1985ag} via a gauge invariant dimension-four operator $\cO_\text{mix} = -\ep F'_{\mu\nu}F^{\mu \nu}/2$, where $F'$ is the field strength of the dark photon $A'$ and $F$ is the field strength of the Standard Model photon $A$. This operator induces a coupling of the dark photon with the electromagnetic current \begin{equation}\label{dpInter}
\cL = \ep A'_\mu J^{\mu}_\text{EM}
\end{equation}
proportional to the kinetic mixing parameter $\ep$. Most searches for the dark photon rely on the observation of the products of its decay after it has been produced in low-energy terrestrial experiments \cite{Batell:2009yf,Essig:2009nc,Reece:2009un,Bossi:2009uw,Bjorken:2009mm,Batell:2009di,Aubert:2009af,Freytsis:2009bh,Essig:2010xa,Essig:2010gu,Merkel:2011ze,Abrahamyan:2011gv,Archilli:2011zc}, although some search strategies are sensitive even to invisibly decaying or effectively stable dark photons \cite{Aubert:2008as, Redondo:2008ec, Nelson:2011sf, An:2013yfc, Chang:2016ntp}.

We are interested in the visible decays of low-mass dark photons. Below the $e^+e^-$ threshold, the dominant decay channel is $A' \to \gamma \gamma \gamma$. [The weak decay $A' \to \nu \bar \nu$ is effectively suppressed by $(m'/m_{W^\pm})^4 \sim 10^{-21} (m'/m_e)^4$, while decay to two photons is forbidden by the Landau-Yang theorem.] The three photon decay width was previously estimated from the Euler-Heisenberg effective action \cite{Pospelov:2008jk}, which we discuss shortly below.  But since the Euler-Heisenberg effective action provides only the leading term in the derivative expansion with a small parameter given by $k_\gamma^2/m_e^2$, the derived decay width is expected to be valid provided $m' \ll m_e$.  For characteristic energies close to the electron mass, as expected for a decaying dark photon of mass in the range $m_e \lesssim m' \lesssim 2m_e$, this approximation breaks down, requiring a more exact treatment.

In this work, we improve the estimate for the $A' \to \gamma \gamma \gamma$ decay width by evaluating it in quantum electrodynamics at one-loop order exactly, obtaining a result  valid for all dark photon masses below the $e^+e^-$ threshold.   Ultimately, we will investigate the bounds for a dark photon that comprises a non-negligible cosmological density in the early universe \cite{Redondo:2008ec}. Interestingly, a small region in parameter space at $\ep \simeq 10^{-4}$ and $850\kev \lesssim m' \leq 2m_e$ is reopened so that a dark photon with these properties is allowed by all model-independent considerations.

\section{Corrections to Euler-Heisenberg Lagrangian(s)}
\begin{figure}[t]
\begin{center}
\includegraphics[width=.28\textwidth]{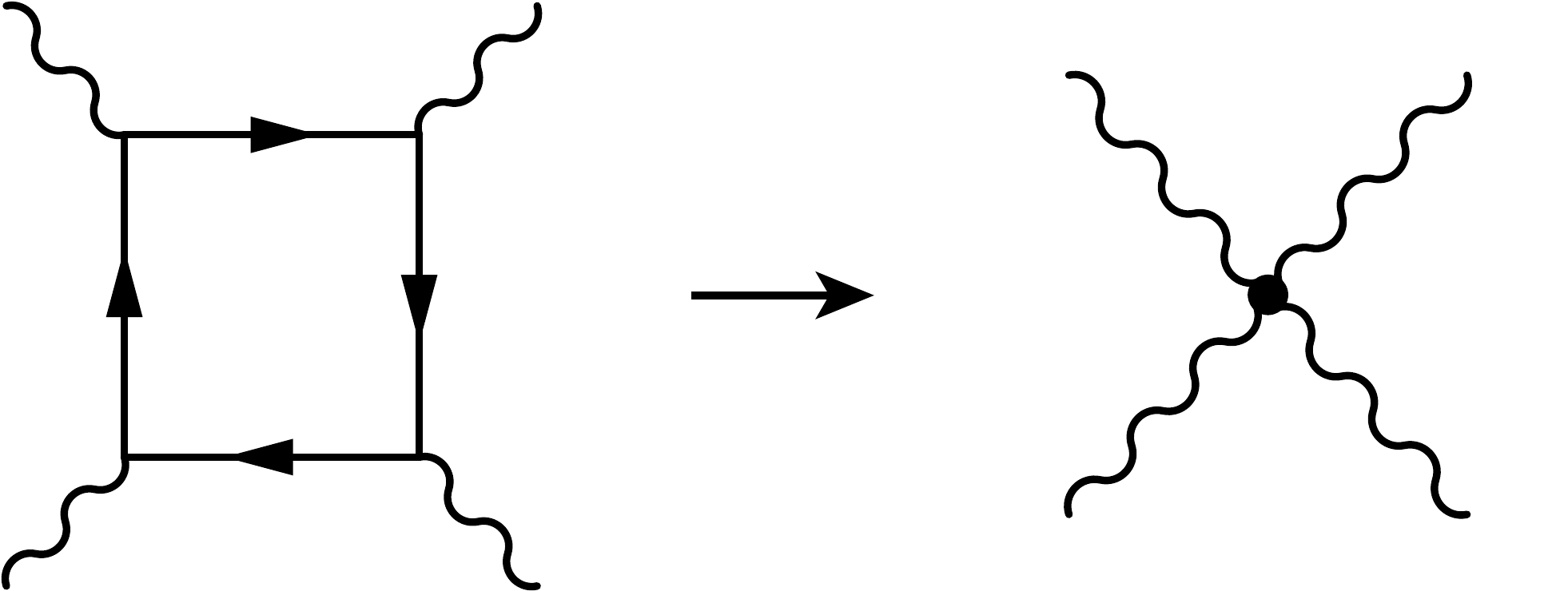}
\caption{The dynamics of an abelian gauge theory below all charged particle masses includes gauge boson self-scattering. At leading order, this is governed by the Euler-Heisenberg Lagrangian.}
\label{fd-EH}
\end{center}
\end{figure}

The Euler-Heisenberg Lagrangian is the low-energy effective Lagrangian that is matched to quantum electrodynamics at energies below the electron mass \cite{Heisenberg:1935qt}. This Lagrangian predicts gauge boson self-couplings that do not exist at higher energies, leading to ``nonlinear'' photon interactions \cite{Costantini:1971cj}. In general, we may construct an Euler-Heisenberg-like Lagrangian for any (product of) gauge theories after the lightest charged particles have been integrated out. Gauge bosons with bifundamental matter acquire any contact interactions that are the remnant of charge-conserving interactions of box diagrams of heavy matter, as shown in \Fig{fd-EH}.

The Euler-Heisenberg Lagrangian contains interactions of arbitrarily many (even) numbers of photons. In four dimensions, the term involving four photons is \cite{Ritz:1995nt}
\beq \label{EH-4photon}
\cL_{\EH4\gamma} = \frac{\alpha^2}{90 m^4}\bL \pL F_{\mu \nu}F^{\mu \nu} \pR^2 + \frac74 \pL F_{\mu \nu} \tilde F^{\mu \nu} \pR^2 \bR .
\eeq
This famously allows the computation of a cross section for light-by-light scattering \cite{Euler:1935zz}. Replacing one of the four photons with a dark photon will give us the Lagrangian for the process that we are interested in. Accounting for the reduced symmetry and inserting a factor $\ep$ for the electron coupling to the dark photon, we find that \Eq{EH-4photon} generalizes to 
\alg{ \label{EH-darkphoton}
\cL_\EH{dark} &= \frac{\ep \alpha^2}{45 m^4} \big( 14 F'_{\mu \nu}F^{\nu \lambda} F_{\lambda \rho} F^{\rho \mu} -\\ & \qquad\qquad\qquad\qquad- 5  F'_{\mu \nu}F^{\mu \nu} F_{\alpha \beta}F^{\alpha \beta} \big),
}
in agreement with \cite{Pospelov:2008jk}.

However, the Lagrangians in \Eqs{EH-4photon}{EH-darkphoton} do not produce the full gauge boson scattering cross sections for arbitrary energies. This is well known for QED: the cross section for light-by-light scattering was obtained by Karplus and Neuman \cite{Karplus:1950zz} and was fully explored by de Tollis and collaborators \cite{Costantini:1971cj}. There, they reproduced the $m_e^{-8}$ scaling predicted by the Euler-Heisenberg Lagrangian at small energy transfer, but found that this deviated from the full result close to the electron mass.

The reason the Euler-Heisenberg Lagrangian fails close to the electron mass is because it is the first term in an expansion in derivatives on the field strength. Although the Euler-Heisenberg Lagrangian is a nonperturbative object that contains infinitely many terms, each term includes $2n$ products of field strength tensors, for integer $n \geq 2$, with $4(n-1)$ powers of $m$ in the denominator of each term. No additional terms with more derivatives appear in this expansion. Thus, the Lagrangians obtained from the Euler-Heisenberg Lagrangian, such as the Lagrangians in \Eqs{EH-4photon}{EH-darkphoton} as well as those with additional photon fields, are only the leading terms of the $2n$ photon interaction in the $m \to \infty$ limit. Each of these leading terms receive $\cO(1)$ corrections when the components of $\pt_\alpha F_{\mu \nu}/m$ are comparable to those of $F_{\mu \nu}$. Higher derivative terms fill out the exact effective theory at energies comparable to the loop particle mass. Infinitely many derivatives are needed to get the exact result, indicating a nonlocality at small distances.

Of course, the full theory that completes the Euler-Heisenberg Lagrangian is known, and this nonlocality is resolved by the electron mass. (Henceforth we will specialize to the QED case, with an electron in the\begin{figure}[h]
\begin{center}
\includegraphics[width=.14\textwidth]{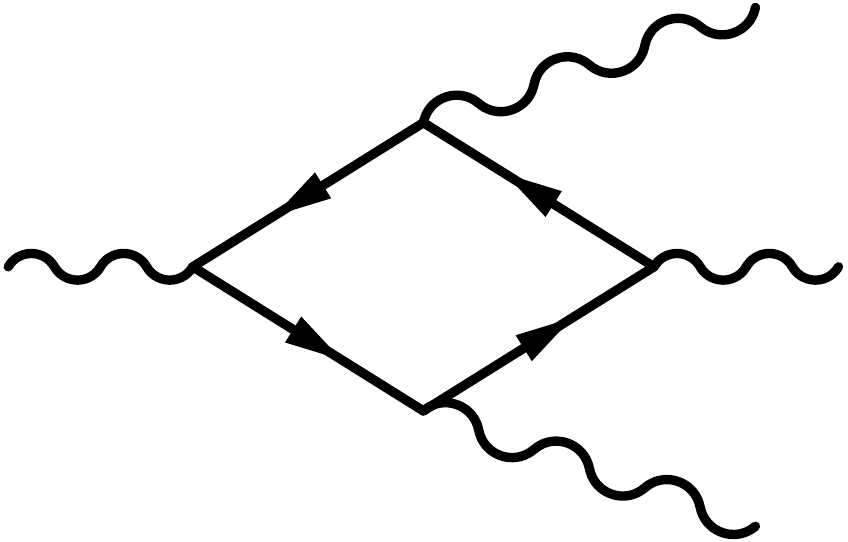}\,\,
\includegraphics[width=.14\textwidth]{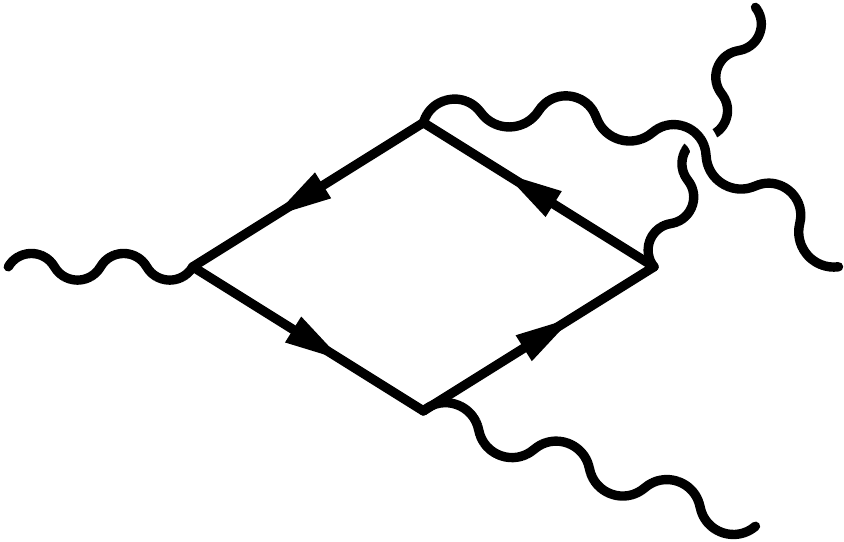}\,\,
\includegraphics[width=.14\textwidth]{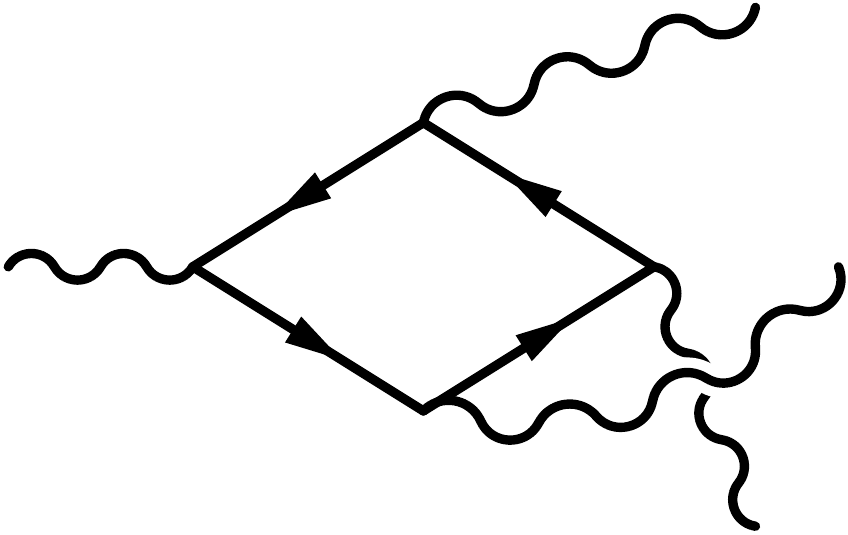}
\caption{Three of six Feynman diagrams contributing to the dark photon decay amplitude at one loop order.  The particle in the loop is an electron.  The three other diagrams are obtained by reversing the electron flow.}
\label{oneloopdiagrams}
\end{center}
\end{figure} loop.) For four-photon interactions, the first two subleading Lagrangian terms containing two and four derivatives have in fact been obtained by a matching calculation using the full QED scattering cross section \cite{Dicus:1997ax}. By calculating in QED, \cite{Dicus:1997ax} found that corrections to light-by-light scattering are an order of magnitude when each incoming photon energy equals $m_e$. The corrections to the three photon decay will na\"ively be larger for a dark photon just below the electron positron pair production threshold: the incoming energy is large compared to $m_e$, while the energy of the outgoing legs is not small compared to $m_e$, since in general $\omega_\gamma \sim \cO(m'/3)$.

 \FloatBarrier

\section{Dark Photon Decay}

We now calculate the dark photon partial width $A'(k') \rightarrow \gamma(k_1) \gamma(k_2) \gamma(k_3)$ exactly at one loop order in quantum electrodynamics.   Three of the six diagrams contributing to the decay amplitude are shown in \Fig{oneloopdiagrams}.  By charge conjugation invariance, the other three diagrams with opposite electron flow are equal in value to these diagrams.  Denoting the amplitude as
\begin{equation}
i\mathcal{M} = i \cA^{\mu\nu\rho\sigma}\, \epsilon(k')_\mu \epsilon^*(k_1)_\nu\epsilon^*(k_2)_\rho \epsilon^*(k_3)_\sigma\,,
\end{equation}
where the $\epsilon(k_i)$ are the polarization four-vectors of the external photons, the Feynman amplitude for the dark photon decay derived from \Eq{dpInter} is
\begin{widetext}
\begin{align}\label{eq:decayampl}
\nonumber i \cA^{\mu\nu\rho\sigma} =& \, 2\epsilon e^4\Bigg[\int\frac{d^4 p}{(2\pi)^4} \frac{\tr [\gamma^\mu(\slashed{p}+m_e)\gamma^\nu(\slashed{p}+\slashed{k}_1+m_e)\gamma^\rho(\slashed{p}+\slashed{k}_1+\slashed{k}_2+m_e)\gamma^\sigma(\slashed{p}+\slashed{k}'+m_e)]}{[p^2 + m_e^2][(p+k_1)^2 + m_e^2][(p+k_1 + k_2)^2 + m_e^2][(p+k')^2 - m_e^2]}\\
\nonumber&+\int\frac{d^4 p}{(2\pi)^4} \frac{\tr [\gamma^\mu(\slashed{p}+m_e)\gamma^\rho(\slashed{p}+\slashed{k}_2+m_e)\gamma^\nu(\slashed{p}+\slashed{k}_1+\slashed{k}_2+m_e)\gamma^\sigma(\slashed{p}+\slashed{k}'+m_e)]}{[p^2 + m_e^2][(p+k_2)^2 + m_e^2][(p+k_1 + k_2)^2 + m_e^2][(p+k')^2 - m_e^2]}\\
 &+\int\frac{d^4 p}{(2\pi)^4} \frac{\tr [\gamma^\mu(\slashed{p}+m_e)\gamma^\nu(\slashed{p}+\slashed{k}_1+m_e)\gamma^\sigma(\slashed{p}+\slashed{k}'-\slashed{k}_2+m_e)\gamma^\rho(\slashed{p}+\slashed{k}'+m_e)]}{[p^2 + m_e^2][(p+k_1)^2 + m_e^2][(p+k' - k_2)^2 + m_e^2][(p+k')^2 - m_e^2]}\Bigg]\,.
\end{align}
\end{widetext}
We evaluate the amplitude and decay width using the algebraic reduction method of Passarino and Veltman \cite{Passarino:1978jh} implemented in {\sc Package-X} \cite{Patel:2015tea}.  %Using energy momentum conservation $k' = k_1 + k_2 + k_3$ to eliminate $k_3$ and transversality of the vector boson polarization tensors to replace
%\begin{align}
%\nonumber k'.\epsilon(k') = 0 &\Rightarrow k'^\mu \rightarrow 0 \\
%\nonumber k_1.\epsilon^*(k_2) = 0 &\Rightarrow k_1^\nu \rightarrow 0 \\
%k_2.\epsilon^*(k_2) = 0 &\Rightarrow k_2^\rho \rightarrow 0\,,
%\end{align}
%the covariant decomposition of the amplitude is written as a linear combination of 138 tensor structures.
We have checked that the total amplitude is free of ultraviolet divergences and that the amplitude is transverse to all external momenta
\begin{equation}
k'_\mu \cA^{\mu\nu\rho\sigma} \! = k_{1\nu} \cA^{\mu\nu\rho\sigma} \!=  k_{2\rho} \cA^{\mu\nu\rho\sigma} \!= k_{3\sigma} \cA^{\mu\nu\rho\sigma} \!= 0\,,
\end{equation}
as required by the abelian Ward identity.  After the algebraic reduction to scalar functions, we square the amplitude, sum over final photon polarizations, and average over the initial dark photon polarization using the completeness relation valid for transverse amplitudes
\alg{
\oL{ \mL \cM \mR }^2 &= \frac{1}{3} \cA^{\mu\nu\rho\sigma}(\cA^*)^{\dot \mu\dot \nu\dot \rho\dot \sigma} \times\\
& \enspace \times \pL -g_{\mu\dot \mu}+ \frac{k'_\mu k'_{\dot \mu}}{m'^2}\pR(-g_{\nu\dot \nu})(-g_{\rho\dot \rho})(-g_{\sigma\dot \sigma})\\
&= \frac{1}{3}\cA^{\mu\nu\rho\sigma}( \cA^*)_{\mu\nu\rho\sigma}\,.
}
We obtain the dark photon decay width by integrating over the three body phase space in terms of the Dalitz variables $m_{12}^2 = (k_1 + k_2)^2$ and $m_{13}^2 = (k_1 + k_3)^2$\,,
\begin{equation} \label{phase-space-integration}
\Gamma_{\gamma\gamma\gamma} =  \frac{1}{(2\pi)^3}\frac{1}{32 m'^3} \frac{1}{6} \int_0^{m'^2}dm^2_{12} \int_0^{m'^2-m_{12}^2} dm^2_{13} \oL{ \mL \cM \mR }^2\,,
\end{equation}
where the factor of $1/6$ accounts for identical final state particles.

\begin{figure}[t]
\begin{center}
\includegraphics[width=.47\textwidth]{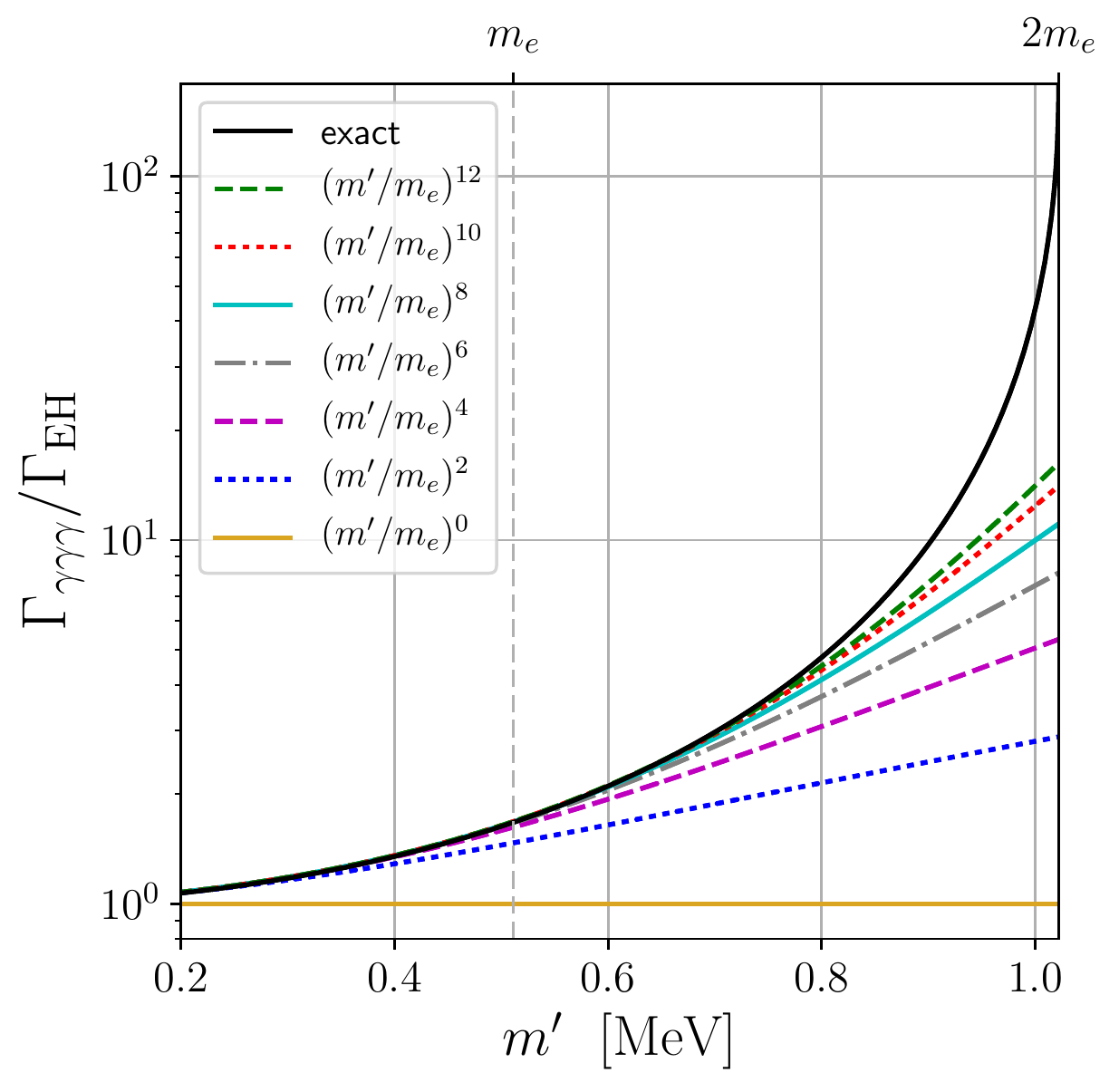}
\caption{Dark photon decay width to three photons in units of the width obtained in the Euler-Heisenberg limit. We show corrections including terms up to sixth order in the variable $m'^2/m_e^2$ as well as the exact correction (in black).}
\label{width-exact}
\end{center}
\end{figure}

There are two potential sources of numerical instability in the numerical evaluation of the decay amplitude. If the dark photon mass is much smaller than the electron mass $m'^2 \ll m_e^2$, or on the boundary of the phase space (where a final state invariant mass becomes small compared to the electron mass, {\it e.g.}~$m_{12}^2, m_{13}^2$ or $m_{23}^2 \ll m_e^2$) the kinematic Gram determinant becomes small and the Passarino-Veltman reduction formula becomes invalid. We circumvent this problem in {\sc Package-X} by evaluating the amplitudes with sufficient working precision to achieve a numerically stable result throughout the entire physical region. With sufficient run time, we can attain any desired precision even arbitrarily close to $m'=2m_e$. We show the exact evaluation of the decay width in \Fig{width-exact}. We provide a supplemental file containing the data points we used to generate the black line.

\begin{table}[b]
\begin{center}
\begin{tabular}{|c|c|c|} \hline
 & $c_k$ & $c_k \times 4^k$\\  \hline
$c_1$ & $\frac{335}{714}$ & 1.88 \\  \hline
$c_2$ & $\frac{128,941}{839,664}$ & 2.46 \\  \hline
$c_3$ & $\frac{44,787}{1,026,256}$ & 2.79 \\  \hline
$c_4$ & $\frac{1,249,649,333}{108,064,756,800}$ & 5.92 \\  \hline
$c_5$ & $\frac{36,494,147}{12,382,420,050}$ & 3.02 \\  \hline
$c_6$ & $\frac{867,635,449}{1,614,300,688,000}$ & 2.20 \\  \hline
\end{tabular}
\end{center}
\caption{Series coefficients of the large electron mass expansion of the dark photon decay $A'\rightarrow\gamma\gamma\gamma$ given in \Eq{width-series}. The final column represents the correction to the decay width due to each term at the threshold value of $m'^2/m_e^2=4$ relative to the Euler-Heisenberg estimate \Eq{wid-EH}.}
\label{dpd-width-coeffs}
\end{table}%

By differentiating the Passarino-Veltman functions arising in the covariant decomposition of the amplitude in \Eq{eq:decayampl}, we may also obtain an analytic representation of the partial width as a series in inverse electron mass:
\beq \label{width-series}
\Gamma_{\gamma\gamma\gamma} = \Gamma_\eh \bL 1 + \sum_{k=1}^\infty c_k \pL \frac{m'^2}{m_e^2} \pR^{k} \bR,
\eeq
which we expect to converge out to the onset of physical threshold at $m'^2/m_e^2 = 4$.  We confirm that the leading term $\Gamma_\text{EH}$ is determined by the amplitude in the Euler-Heisenberg limit derived from \Eq{EH-darkphoton},
\beq \label{wid-EH}
\Gamma_\eh = \frac{17 \ep^2 \alpha_{\rm EM}^4}{11664000\, \pi^3} \frac{m'^9}{m_e^8} \simeq 1 \s^{-1} \pL \frac\ep{0.003} \pR^2 \pL \frac{m'}{m_e} \pR^9,
\eeq
in agreement with \cite{Pospelov:2008jk}.  In the context of the Euler-Heinsberg effective theory discussed in the previous section, the remaining terms in \Eq{width-series} arise from corrections to the Euler-Heisenberg effective action involving derivatives of the field strength tensors. We give the first six coefficients $c_k$ in \Tab{dpd-width-coeffs}, and in \Fig{width-exact} we plot the corresponding corrections to the partial width. 

As the dark photon mass approaches the electron pair production threshold, each term supplies a correction that is individually larger than the width in the Euler-Heisenberg limit. We demonstrate this in the third column of \Tab{dpd-width-coeffs}. Thus, although the sixth order correction gives an acceptable approximation to the exact width up to $m' \simeq 800\kev$, as shown in \Fig{width-exact}, asymptotically many terms are required for an accurate estimate of the partial width for $m' \simeq 2m_e$. At threshold, we find that the dark photon partial width is over two orders of magnitude larger than the width in the Euler-Heisenberg limit. This means that the dark photon lifetime is  drastically shorter in this mass range than previously appreciated. This has important phenomenological consequences, which we discuss next.

\section{Updated constraints on Dark Photon Mass Below $2m_e$}

\begin{figure*}[htp]
\begin{center}
\includegraphics[width=.7\textwidth]{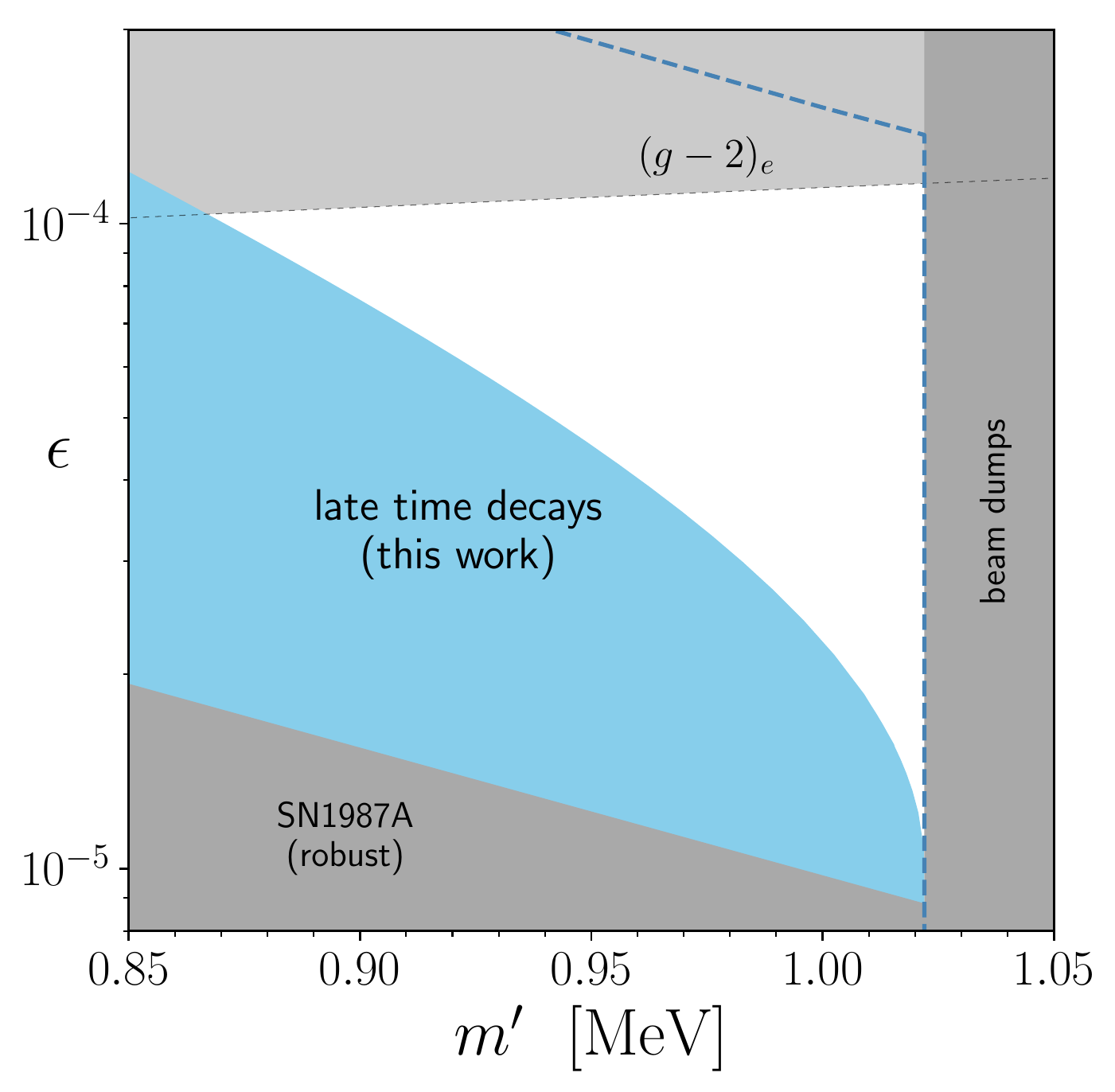}
\caption{Constraints on the dark photon parameter space. In the blue region the lifetime of the dark photon to decay to three photons exceeds one second. There is an open a window near $10^{-5} \lesssim \ep \lesssim 10^{-4}$ and $850 \kev \lesssim m' \lesssim 1 \mev$. We also show constraints from SN1987A using the ``robust'' results from \cite{Chang:2016ntp}, constraints from the anomalous magnetic moment of the electron \cite{Pospelov:2008zw}, constraints from beam dump experiments \cite{Essig:2013lka}, and the results from the lifetime only using the leading-order Euler-Heisenberg-like width (dashed blue).}
\label{dpd-constraint-plot}
\end{center}
\end{figure*}

A dark photon with kinetic mixing parameter larger than around $\ep \gtrsim 10^{-12}$ can be in thermal equilibrium at some point in the early universe, and can thus play the role of cosmological dark matter or otherwise affect the evolution of the early universe \cite{Redondo:2008ec, Nelson:2011sf}. If the dark photon mass satisfies $m' \leq 2m_e$, the dark photon is quite long-lived because the decay to three photons is loop-mediated, contains four factors of $\alpha_{\rm EM}$, and must go into three body phase space, as in \Eqs{phase-space-integration}{wid-EH}. Thus, dark photons with mass $m' \leq 2m_e$ generically have long lifetimes, modulated by the kinetic mixing factor $\ep$. The decay to three photons can happen late enough to alter Standard Model physics in epochs where the thermal history of the universe is well established. In this section, we describe the effects of such a long-lived dark photon, and we detail the bounds from considerations of this kind of late-decaying particle.

Dark photon decays should not occur between the time of heavy element nucleosynthesis (BBN) and the formation of the cosmic microwave background radiation (CMB), because the additional entropy injection would conflict with the observed agreement between the baryon to photon ratio at BBN and the CMB. Decays from the time of formation of CMB through the cosmological dark ages are ruled out by limits on the $\mu$ and $y$ distortions of the CMB blackbody spectrum. Decays after the epoch of reionization until the present time are forbidden by limits on the amount of diffuse extragalactic background light. For $m' \leq 2m_e$, we can then rule out lifetimes
\beq
1 \s  \leq \Gamma_{\gamma\gamma\gamma} \leq T_{\rm Univ},
\eeq
where $T_{\rm Univ}$ is the age of the universe. The bounds on the background light are conservatively obtained from saturating observations of the high-latitude flux of X-ray photons, as in \cite{Essig:2013goa}. We illustrate the ramifications for the dark photon parameter space in \Fig{dpd-constraint-plot}.

We mention that in this work we neglect possible bounds from degree of freedom counting in the early universe. For the mixing angles considered here, the dark photon will be resonantly produced in the early universe and may remain in kinetic equilibrium until after the time of neutrino decoupling \cite{Redondo:2008ec}, potentially adding to the radiation budget of the universe. This could speed up the expansion rate and adversely affect sensitive BBN processes. However, corresponding negative contributions to the radiation budget could come from slightly heavier particles coupled to electric charge \cite{Boehm:2013jpa} (which plausibly exist in a complete model of the dark sector) or from the post-BBN decay of the dark photon. This renders the counting of additional degrees of freedom model dependent for the scenario considered. Finding a UV-complete model necessary to avoid all of the bounds discussed here is a compelling goal for future work.

We point out an open window free of all constraints where $10^{-5}\lesssim \ep \lesssim 10^{-4}$ and $850 \kev \lesssim m' \leq 2m_e$. This is precisely the mass range where subleading corrections to the Euler-Heisenberg-like Lagrangian are large. Our exact computation of the decay width is crucial to revealing this window: using the leading order Euler-Heisenberg-like Lagrangian, one would conclude that dark photons of this mass have lifetimes longer than one second, leading to entropy injection after BBN and ruling out all mixing angles below $\ep \lesssim 0.003\times(m'/m_e)^{-9/2}$ \cite{Chang:2016ntp}. However, because threshold corrections increase the width by up to two orders of magnitude, we find that dark photons in this mass range decay well before BBN.

This window remains open because a dark photon in this window is {\it weakly} coupled enough that it does not affect the anomalous magnetic moment of the electron \cite{Pospelov:2008zw, Essig:2013vha} but {\it strongly} coupled enough that it is effectively trapped inside the proto-neutron star during the detonation of Supernova 1987A: using the most conservative models of the core collapse process, we find that a dark photon in this parameter space does not carry out more energy than is released in Standard Model neutrinos \cite{Chang:2016ntp}. About half of the parameter space is ruled out assuming the ``fiducial'' analytical profile, however \cite{Chang:2016ntp}.

This range is potentially constrained by considerations of the SLAC millicharge experiment \cite{zoya}, which is sensitive to any light dark matter that can scatter off nuclei. In the near future, this window will also be subject to searches performed by new beam dump experiments that will be sensitive to any particle that can escape the beam as missing energy \cite{Battaglieri:2014qoa, ldmx}. As a complementary technique, it is possible that improvements of the measurement of the anomalous magnetic moment of the electron will soon be sufficiently improved to constrain this window of parameter space \cite{gardner}. Thus, this exciting window for new physics will be comprehensively and complementarily probed in a variety of upcoming experiments.

\section*{Acknowledgments}
We thank Zackaria Chacko, Rouven Essig, Marat Freytsis, Susan Gardner, and Jakub Scholtz for discussions. SDM is supported by NSF-PHY-1316617. H.R. is supported in part by NSF CAREER award NSF-PHY-1056833 and NSF award NSF-PHY-1620628.

\bibliography{dpd-eh}

\end{document}